\documentclass[aps,prl,reprint,amsmath,amsfonts,superscriptaddress]{revtex4-1}%reprint: two column, linenumbers

\usepackage[dvipsnames]{xcolor}

\usepackage[colorlinks,allcolors={ForestGreen}]{hyperref}
\usepackage{graphicx}% Include figure files
\usepackage{dcolumn}% Align table columns on decimal point
\usepackage{bm}% bold math
\usepackage[T1]{fontenc}
\usepackage{color}
\def\sat{NNLO$_{\text{sat}}$}

\begin{document}

%\preprint{APS/123-QED}

\title{Lepton scattering from $^{40}$Ar and $^{48}$Ti in the quasielastic peak region} %\\with Forced Linebreak}% Force line breaks with \\
%\title{First principle computations of lepton scattering on $^{40}$Ar} %\\with Forced Linebreak}% Force line breaks with \\
%\thanks{A footnote to the article title}%

%####################### first author #####################################################

\author{C.~Barbieri}
\email{C.Barbieri@surrey.ac.uk} % displayed before bib
\affiliation{Department of Physics, University of Surrey, Guilford GU2 7XH, United Kingdom}
\author{N.~Rocco}
\affiliation{Fermi National Accelerator Laboratory, Batavia, IL 60510, USA}
\affiliation{Physics Division, Argonne National Laboratory, Argonne, IL 60439, USA}
\affiliation{Department of Physics, University of Surrey, Guilford GU2 7XH, United Kingdom}
\author{V. Som\`a}
\affiliation{IRFU, CEA, Universit\'e Paris-Saclay, 91191 Gif-sur-Yvette, France}

%####################### second author #####################################################

%\author{Charlie Author}
%\homepage{http://www.Second.institution.edu/~Charlie.Author}
%\affiliation{Second institution and/or address\\ This line break forced% with \\}
%\affiliation{Third institution, the second for Charlie Author}

%####################### third author #####################################################

%\author{Delta Author}
%\affiliation{Authors' institution and/or address\\ This line break forced with \textbackslash\textbackslash}
%\collaboration{CLEO Collaboration}%\noaffiliation 

\date{\today}

\begin{abstract}
Neutron and proton spectral functions of $^{40}$Ar, $^{40}$Ca,  and $^{48}$Ti isotopes are computed using the \hbox{\textit{ab initio}} self-consistent Green's function approach. % and a saturating nuclear Hamiltonian from chiral effective field theory.
The resulting radii and charge distributions are in good agreement with available experimental data. 
The spectral functions of Ar and Ti are then utilized to calculate inclusive ($e$,$e$') cross sections within a factorization scheme and are found to correctly reproduce the recent Jefferson Lab measurements. 
Based on these successful agreements, the weak charged and neutral current double-differential cross sections for neutrino-$^{40}$Ar scattering are predicted in the quasielastic region. 
%quasielastic contribution to neutrino scattering on $^{40}$Ar at 1 GeV. % in the spectral function formalism. 
Results obtained by replacing the (experimentally inaccessible) neutron spectral distribution of $^{40}$Ar with the (experimentally accessible) proton distribution of $^{48}$Ti are compared and the accuracy of this approximation is assessed.

%\begin{description}
%\item[PACS numbers] xxxxxxxxxxx
%\item[DOI] xxxxxxxxxx 
%\end{description}
\end{abstract}

\pacs{Valid PACS appear here}% PACS, the Physics and Astronomy Classification Scheme.
%\keywords{Suggested keywords}%Use showkeys class option if keyword, display desired

\maketitle

%\tableofcontents

\emph{Introduction.} 
Neutrinos are among the most elusive particles in the universe.  They come in three known leptonic flavours, each with an almost zero mass, and they interact with matter weakly. In spite of this, they play relevant roles in extreme astrophysical scenarios such as supernova explosions~\cite{ShenRPC2013nuPNM}. 
Since the discovery of neutrino oscillations, about two decades ago, these particles have been playing a key role in the search of physics beyond the Standard Model. The two most compelling open questions concern the correct hierarchy among the three mass eigenstates and whether the neutrino is its own antiparticle and can be described by a Majorana field~\cite{McDonaldRMP2016,AvignoneRMP2008}. 
The existence of a fourth (sterile) neutrino has also been proposed and could explain the excess of electron neutrinos  from charged current quasielastic (QE) events reported by the MiniBooNe collaboration~\cite{Aguilar2018nue}.

 The new generation of neutrino experiments, such as the short-~\cite{Antonello2015sbn} and long-baseline neutrino~\cite{Abi2018dune} programs will aim at addressing these fundamental questions. 
 In particular, the Deep Underground Neutrino Experiment (DUNE) has the ambitious goal of resolving the hierarchy of mass eigenstates and test for leptonic charge-parity violations.
 These experiments will utilize liquid-argon time-projection chamber technology, which exploits scattering of neutrinos off $^{40}$Ar nuclei contained in the detectors.  In a typical event, one or several hadrons are emitted and detected to reconstruct the flavor and energy of the incident neutrino. 
If the latter is not reconstructed with sufficient accuracy, it is not possible to pin down the oscillation parameters to the precision needed for extracting information on the mass hierarchy~\cite{Acciarri2015dune}.

%For the typical scattering energies of oscillation experiments, up to a few GeV, the impulse approximation (IA) can be safely applied. Within this scheme, the neutrino interacts directly with a nucleon (or a virtual pion) inside the target and the effects of many-body nuclear correlations are embedded in the one- and two-hole spectral functions~\cite{PaviaBook}. On the other hand, modeling the neutrino nucleus interactions for the few GeV region remains %
Modeling neutrino-nucleus interactions in the region of interest for neutrino oscillation experiments, extending up to few GeV, is a very complicated problem~\cite{Katori2017}. 
First, different reaction mechanisms are at play. Depending on the energy transferred by the probe, cross sections are dominated by one- and two-nucleon emission processes in the quasielastic region, excitation of nuclear resonances that subsequently decay into pions and deep inelastic effects leading to hadron production.
Second, a realistic description of nuclear dynamics accounting for many-body correlations in the target is needed. In fact, early models based on a Fermi gas do not convey realistic details of the energy-momentum distributions of the struck nucleon and have proven to be inadequate to reproduce neutrino scattering data~\cite{Aguilar:2008,AguilarArevalo:2010zc}.
Third, electroweak current operators and reaction models need to be validated for the GeV energies at play. Electron scattering data are extremely important to this purpose since they can probe the vector current operators for monochromatic incident beams in a variety of kinematical regions. 
Addressing these points is extremely important for the success of neutrino programs. 
In this regards, very promising results have been obtained combining the impulse approximation (IA) with a realistic spectral function that embeds many-body nuclear correlation effects. This formalism has been extensively tested in the electromagnetic sector and recently generalized to include one- and two-body current processes for both electron- and neutrino-nucleus scattering processes~\cite{Ankowski:2014yfa,Rocco2016prl,Rocco2018escatt,Rocco20192bcurr}.   

The E12-14-012 experiment at Jefferson Lab Hall A recently analyzed the inclusive and exclusive electron scattering on $^{12}$C, $^{40}$Ar and natural Ti targets 
at a fixed beam energy and scattering angle~\cite{Dai2018Tiee1,Dai2019Ar40ee1}. The final goal of this experiment is to study
the properties of the argon nucleus and extract its proton and neutron spectral functions. 
%Unfortunately, electrons do not interact sensibly with neutrons and the latter cannot be efficiently measured in the detectors.
 However, such measurements are typically limited to ejected protons---from ($e$,$e$'$p$) reactions---since neutrons have weaker longitudinal cross sections with electrons and they would also be detected with poorer efficiencies than protons.
For this reason, based on the observation that the neutron spectrum of $^{40}$Ar is mirrored by the proton spectrum of 
Ti isotopes, titanium data have been used to gain indirect information on the neutron spectral function of argon.

In this work we show the results obtained using the spectral functions of Ar and Ti computed within a state-of-the-art \emph{ab initio} theory. 
In order to tackle these open shell nuclei, the Self Consistent Green's Function (SCGF) formalism has been recently generalized in the frame of Gorkov's theory. 
The SCGF is a polynomially-scaling many-body method that allows to efficiently describe nuclei with mass number up to A{\,}$\approx${\,}100.
Using the accurate predictions obtained for proton and neutron spectral functions of $^{40}$Ar and $^{48}$Ti , we calculate quasielastic electron scattering cross section and validate against the JLab experiment to assess their quality. We then compare the theoretical neutron spectral distribution of $^{40}$Ar with the protons in $^{48}$Ti  to quantify the accuracy of the isospin symmetry assumption and provide predictions for neutrino-Ar scattering at the energies relevant to DUNE.  We find that modeling neutrons in Ar upon the proton distribution in Ti is a well justified approximation, once that the relative shifts in the two energy spectra are taken into account.

\emph{Theory.} 
The double differential cross section for inclusive lepton-nucleus scattering can be written as~\cite{PaviaBook}:
\begin{align}
\Big(\frac{d\sigma}{dE^\prime d\Omega^\prime}\Big)_\ell &= C_\ell\; \frac{E_k^\prime}{E_k} \; L_{\mu \nu}W^{\mu \nu} \, ,
\label{eq:cross_sec}
\end{align}
where $L_{\mu \nu}$ is the leptonic tensor and  $k=(E_k,{\bf k})$ and $k^\prime=(E_k^\prime , {\bf k}^\prime)$ are the laboratory four-momenta of the incoming and outgoing leptons, respectively.
The factor $C_\ell=\alpha/(k-k^\prime)^4$ for electrons and \hbox{$C_\ell=G/8\pi^2$} for neutrinos, where $G=G_F$ for neutral current (NC) and $G=G_F \cos\theta_c$ for charged current (CC) processes. The electroweak coupling constants are \hbox{$\alpha\simeq 1/137$},  $G_F=1.1803 \times 10^{-5}\,\rm GeV^{-2}$~\cite{Herczeg:1999} and~\hbox{$\cos\theta_c=0.97425$~\cite{Nakamura2010PDG}}.

The hadron tensor $W^{\mu \nu}$ encodes the transition matrix elements from the target ground state $|\Psi^A_0\rangle$ to the final states $|\Psi^A_f\rangle$ due to the hadronic currents, which include additional axial terms for neutrino scattering.  
For the case of quasielastic  processes at moderate values of the momentum transfer ($|{\bf q}|\gtrsim$ 500 MeV), the impulse approximation 
allows to factorize  $|\Psi^A_f\rangle\rightarrow|{\bf p}^\prime\rangle \otimes |\Psi^{A-1}_n\rangle$ into the outgoing nucleon of momentum ${\bf p}^\prime$ and the residual nucleus in a state $|\Psi^{A-1}_n\rangle$. This leads to~\cite{Rocco2018escatt,Rocco20192bcurr}:
\begin{align}
&W^{\mu\nu}_{\rm 1b}({\bf q},\omega)=\int \frac{d^3 {\bf p}^\prime \; dE}{(2\pi)^3} \frac{m_N^2}{e({\bf p}^\prime)e({\bf p^\prime\!-\!q})} \delta(\omega+E-e(\mathbf{\bf p}^\prime)) \nonumber \\
&\quad \times  \sum_{s}\,   S^h_s({\bf p}^\prime\!-\!{\bf q},E)  \langle p^\prime | {j_{s}^\mu}^\dagger |p^\prime\!-\!q \rangle \langle p^\prime\!-\!q |  j_{s}^\nu | p^\prime \rangle \, ,
\label{had:tens}
\end{align}
where $\omega$ represents the energy transfer, $m_N$ is the nucleon mass, $e({\bf p})$ the energy of a nucleon with momentum ${\bf p}$, the one-body current operators ${j}^\mu_s$ depend on the  spin-isospin degrees of freedom $s$ and $S^h_s({\bf p}, E)$ is the one-hole spectral function normalized to the total number of nucleons.
For two-body currents and hadron production, Eq.~\eqref{had:tens} extends non trivially in terms of one- and two-body spectral functions~\cite{Giusti2005,Barbieri2004O16epp,Barbieri2006NPB,Rocco20192bcurr}.

Final-state interactions (FSI) of the struck nucleon can be accounted for using Glauber  theory~\cite{Benhar1991prc,Barbieri2004prcResc,Barbieri2005PLB,Barbieri2006NPB,Benhar2013prc}.
For the inclusive processes discussed here we follow Ref.~\cite{Benhar2013prc}:
\begin{align}
d\sigma_{FSI}(\omega) &=
 \int d\omega'  \; f_{\bf q}(\omega - \omega' - U_V ) d\sigma(\omega') \; ,
\label{eq:FSI}
\end{align}
where $U_V$ and the function $f_{\bf q}(\omega)$ account for the shift in the cross section and the redistribution of strength away from the quasielastic peak due to interactions 
of the ejected nucleon with the mean field of the residual system and rescattering processes, respectively~\cite{Benhar2005prd,Benhar2013prc}.
Since, to the best of our knowledge, optical potentials for Ar and Ti are not available in the literature, in the present work we use the one of $^{40}$Ca taken from Ref.~\cite{Cooper:1987uy} and the folding function of Ref.~\cite{Benhar2013prc}. 

The internal structure of the target is encoded in the diagonal part of the one-hole spectral function,
\begin{align}
S^h_s({\bf p}, E) &\!= \!\!\sum_n \! \left| \langle \Psi^{A-1}_n | c_s({\bf p}) | \Psi^A_0\rangle \right|^2 \! \delta (E \!- \!E^A_0 \!+\! E^{A-1}_n) ,
\label{eq:Sh}
\end{align}
where 
%$|\Psi^{A-1}_n \rangle$ are the final states of the residual nucleus and 
$c_s({\bf p})$ annihilates a nucleon with momentum ${\bf p}$ and spin-isospin degrees of freedom $s$.
For open-shell nuclei, such as Ar and Ti isotopes, we extract the spectral function from the imaginary part of the normal one-body propagator, $S^h_s({\bf p}, E)=\frac{-1}{\pi}\operatorname{Im}\{G^h({\bf p},{\bf p}; \mu-E)\}$, computed in {\em ab initio} Gorkov self-consistent Green's function (GGF) theory~\cite{Soma2011GkvI,Soma2013GkvRapid,Soma2014GkvII}.  The Gorkov formulation of propagator theory breaks particle-number conservation explicitly and uses a grand canonical Hamiltonian, $\hat\Omega=\hat{H} - \mu_p\hat{Z} - \mu_n\hat{N}$, with chemical potentials $\mu_{p,n}$ tuned to recover the correct number of protons and neutron on average. Breaking of the particle-number symmetry implies the appearance of both normal and anomalous one-body propagators, however, it accounts for pairing correlations and lifts the degeneracies that would otherwise prevent microscopic calculations for open-shell systems.
In GGF theory, the propagator is obtained as solution of Gorkov equations, which generalize standard Dyson equation and encode the many-body expansion in normal and anomalous self-energy terms~\cite{Soma2011GkvI}.

\begin{figure}[t]
\includegraphics[width=0.47\textwidth]{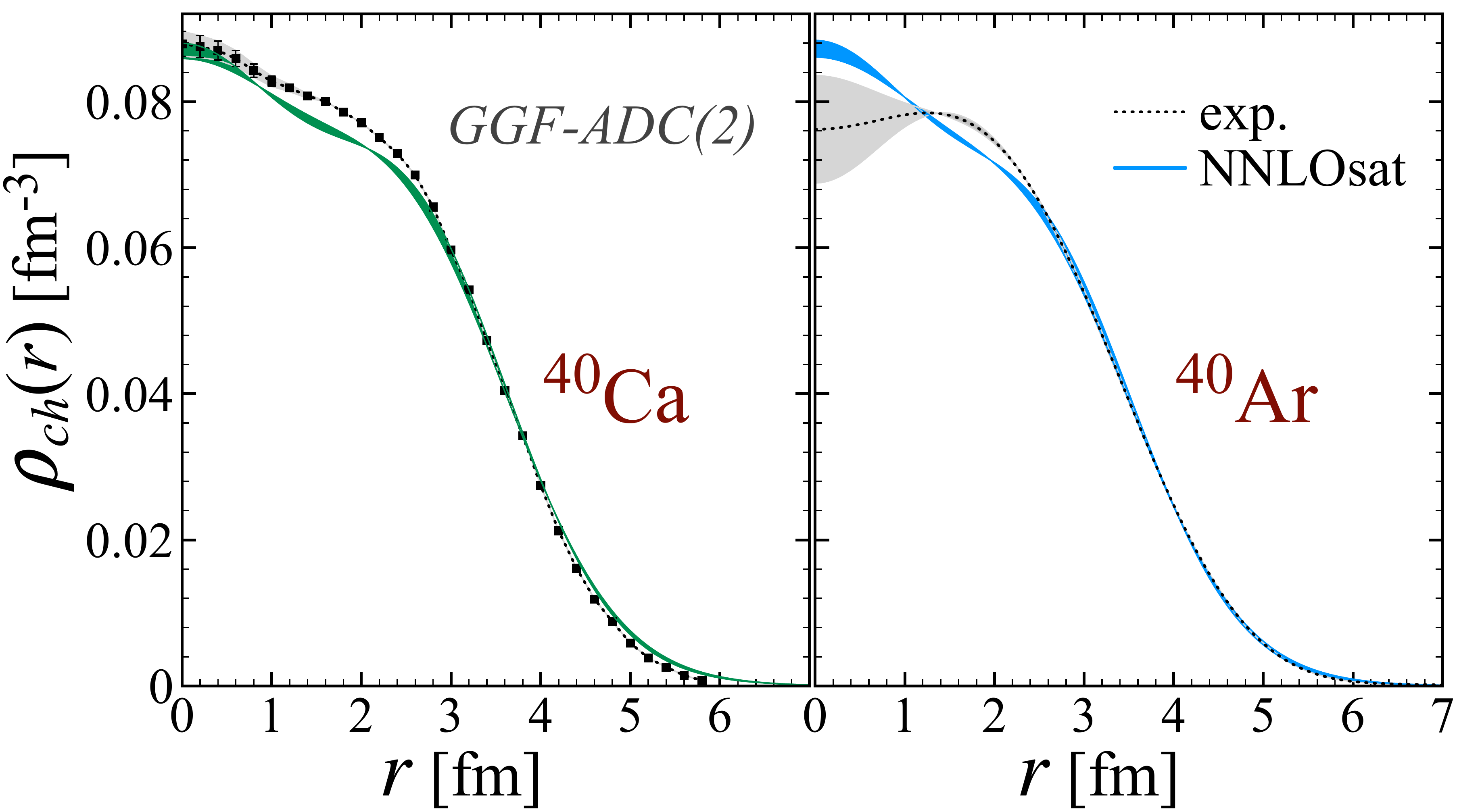}
\caption{Charge density distributions of $^{40}$Ca (left) and $^{40}$Ar (right). Results obtained with the \sat{} interaction in the GGF-ADC(2) approach are compared to experimental data (dotted lines and square points) from Refs~\cite{Emrich83, Ottermann1982Ar40ee1}. The shaded grey areas represent the total experimental error, while the coloured bands displays the theoretical uncertainties  due to model-space convergence.
 \label{fig:rho_ch}}
\end{figure}

\emph{Results.} 
In this work, we solve Gorkov equations using 14 major harmonic oscillators shells and vary the frequency, $\hbar\Omega$, to study the uncertainties resulting from the truncation of the model space.
The self-energy is expanded up to second-order in an optimised reference state (OpRS) propagator (see Refs~\cite{Rocco2018escatt,Soma2019LNL} for details). This many-body truncation, normally referred to as ADC(2), includes triplets of non interacting Gorkov quasiparticles and it incorporates the two hole-one particle (2h1p) configurations of the residual nucleus $|\Psi^{(A-1)}_n \rangle$ that lead to the 2p2h contributions to the final state $| \Psi^A_f \rangle$. Since lepton scattering is sensitive to matter and momentum distribution of the target, we employ the \sat{} chiral interaction of Ref.~\cite{Ekstrom2015nnlosat} that has been shown to reproduce accurately electron scattering on $^{16}$O~\cite{Rocco2018escatt} 
as well as the radii and charge density distributions for isotopes up to $^{48}$Ca~\cite{Hagen2016nature,Lapoux2016prl,Lecluse2017prc,Ruiz2016nature}.
From the analysis of Ref.~\cite{Soma2019LNL}, it is known that the range $\hbar\Omega=14-20$~MeV includes the optimal values for the convergence of both radii and energies. Thus, we perform computations at the extremes of this interval and take the differences in our results as conservative estimates for the theoretical errors due to model space convergence.
The quality of our predictions is demonstrated by Fig.~\ref{fig:rho_ch}, where we compare charge density profiles computed in GGF-ADC(2) for $^{40}$Ca and $^{40}$Ar to experimental data from Refs.~\cite{Emrich83, Ottermann1982Ar40ee1}. 
The resulting charge radii are 3.43(3), 3.52(4) and 3.60(4)~fm
for $^{40}$Ar, $^{40}$Ca  and $^{48}$Ti, respectively, to be compared to the experimental values of  3.427, 3.477 and  3.607~fm~\cite{Angeli13}.  
For $^{40}$Ar we  find point proton and neutron radii of $r_p=3.33(3)$ and $r_n=3.41(4)$~fm,
corresponding to a neutron skin thickness of 0.08(1)~fm.  This value is consistent with the estimate of Ref.~\cite{Payne19}, although their proton and neutron radii computed with the same \sat{} interaction slightly differ from our results.
For $^{40}$Ca, we also performed computations with the more accurate ADC(3) truncation and found negligible changes in the charge density profile. Hence, Fig.~\ref{fig:rho_ch} is substantially converged with respect to many-body truncations, as already found in Ref.~\cite{Lecluse2017prc} for 
S and Si isotopes.

\begin{figure}[t!]
\includegraphics[width=.5\textwidth]{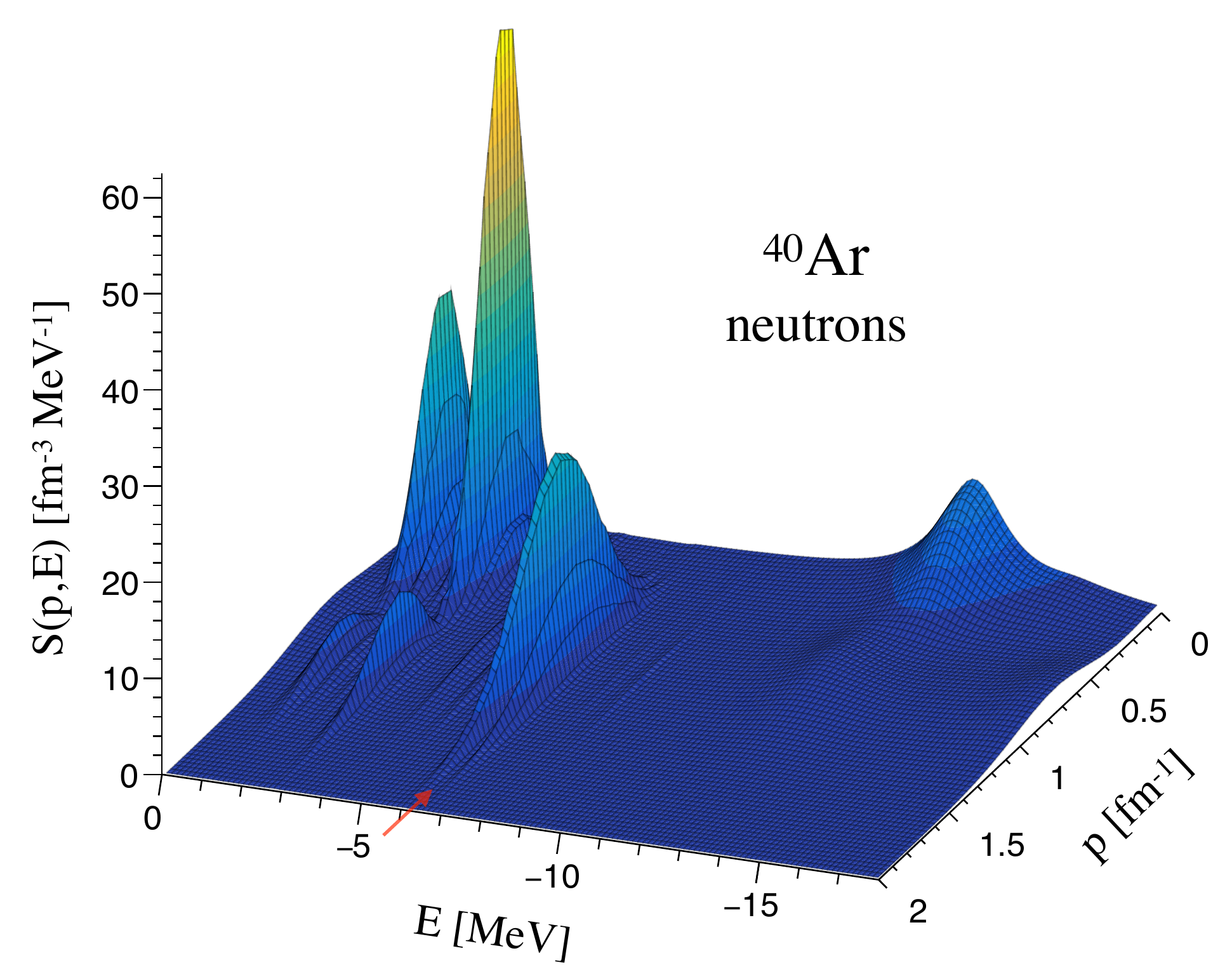}
\caption{Neutron spectral function of $^{40}$Ar computed from GGF-ADC(2) using the \sat{} chiral Hamiltonian.
 Particle and hole spectral functions are identified respectively above and below the Fermi energy situated at $-6.3$~MeV (red arrow). \label{fig:Sh_Ar40}}
\end{figure}

The key point in the factorization approach to the hadronic contributions of Eq.~\eqref{had:tens} is that, within the limit of validity of the IA, the scattering process can be described as an incoherent sum of lepton scattering amplitudes on bound nucleons, provided that the process is averaged over the probability of finding nucleons in the target with given initial momentum and energies.  The hole spectral function, $S^h_s({\bf p}, E)$ encodes exactly this information, since it has a specific interpretation as the joint probability of removing a nucleon with momentum ${\bf p}$ after transferring energy $E$ to the target nucleus.
As an example, Fig.~\ref{fig:Sh_Ar40} displays the computed $S^h_s({\bf p}, E)$ for neutron removal from $^{40}$Ar, as well as the corresponding neutron-addition part $S^p_s({\bf p}, E)$.
The dominant peaks at low separation energies (small values of $|E|$) carry information on the momentum distribution of nucleons occupying the valence `orbits' near the Fermi surface. As the separation energy increases, the distribution becomes more spread and covers the particles associated with the nuclear core. For large separation energies (not shown here), $E<$~-60~MeV and  $|{\bf p}|>$~2~fm$^{-1}$, the spectral function presents a mild tail carrying the strength at larger momenta, typically associated with short-distance interactions among nucleons. The correlation between high missing energies and momenta in such tail is a very general feature for self-bound systems (such as nuclei) and it is dictated by kinematical constraints.  
It must be stressed that the amount of spectral strength in this tail depends on the scale resolution associated to the chosen nuclear Hamiltonian. In spite of being a relatively soft interaction, with a cutoff of 450~MeV/c, \sat{} still predicts the presence of larger momentum components. Nevertheless, such components are clearly weaker than the ones obtained from high-accuracy (and high-cutoff) phenomenological forces such as AV18~\cite{Rocco2018escatt}.

\begin{figure}[t]
\includegraphics[width=0.47\textwidth]{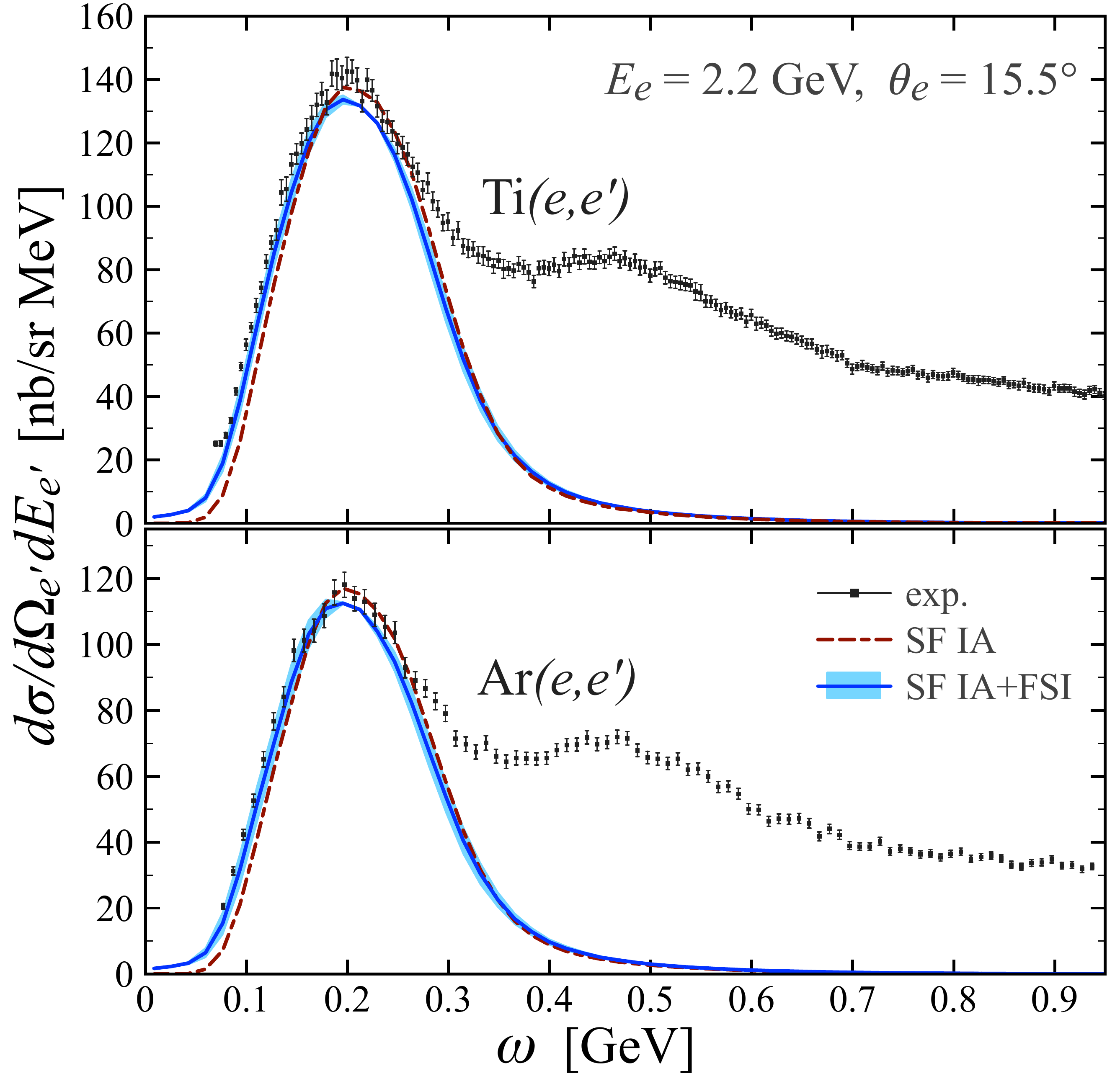}
\caption{Inclusive Ti($e$,$e$') (top) and Ar($e$,$e$') (bottom) cross sections at 2.2~GeV and 15.5${}^\circ$ scattering angle. The solid (dashed) line shows the quasielastic cross section with (without) the inclusion of FSI. 
For the FSI results, the theoretical uncertainties coming from model-space convergence are also shown as a shaded band.
Experimental data are taken from Ref.~\cite{Dai2018Tiee1,Dai2019Ar40ee1} and show both the quasielastic peak and the contribution from meson production at larger missing energies. \label{fig:jlab_xsec}}
\end{figure}

Many-body correlations control the location of the hole spectral strength. Since it remains mostly contained in a region of approximately $-100~\text{MeV}<E<0~\text{MeV}$, the very fine details of the distribution
 are less important for the inclusive reactions and the lepton probes at few GeV energies that are relevant for the present work. 
In fact, models based on the relativistic Green's function approach~\cite{Meucci2003RGF,Meucci2014RHB} describe well the 
electron scattering data in the quasielastic peak starting from available relativistic mean-fields and optical potentials~\cite{Dai2019Ar40ee1}.
This description is already superior to a Fermi-gas model, to an extent that the description of neutrino-nucleus scattering is noticeably impacted~\cite{Benhar:2009wi,Benhar:2015wva,Rocco20192bcurr}.
The \textit{ab initio} spectral function of Fig.~\ref{fig:Sh_Ar40} is computed directly from the underlying two- and three -nucleon interactions and it contains even more detailed information about the structure of the nuclear target.  The knowledge of $S^h_s({\bf p}, E)$ can then impact the accurate determination of the cross sections, especially for exclusive events.
The quality of our spectral functions is tested by computing the inclusive electron scattering on $^{40}$Ar and $^{48}$Ti at the energy and kinematics of the E12-14-012 JLab experiment. The resulting cross sections are displayed in Fig.~\ref{fig:jlab_xsec} as a function of the energy transfer and reproduce closely the quasielastic peak from experimental data. In the present calculation, we have neglected two-nucleon currents and meson-production contributions that dominate the cross section at higher energy transfer~\cite{Rocco2016prl}.  The dashed and solid curves in the figures demonstrate the effect of FSI.
 Note that the colored band in the FSI curve also shows the uncertainty from model space convergence that has been estimated as discussed above.  This is representative of both curves and shows that our calculations are near full convergence with respect to the model space.
The inclusion of FSI produces a small shift in the position of the quasielastic peak that improves the description for \hbox{$\omega <$ 180~MeV}. On the other hand, strength is removed from the maximum of the peak and moved to the tail.  Hence, the prediction based on the \sat{} interaction and GGF-ADC(2) for ground state correlations slightly underestimates the experimental data at the peak.
Overall, the discrepancy is still rather small and it is compatible with the larger uncertainties that are intrinsic with the accuracy of state-of-the-art nuclear forces~\cite{Hebeler2015AnnRev}.
\begin{figure}
\includegraphics[width=0.46\textwidth]{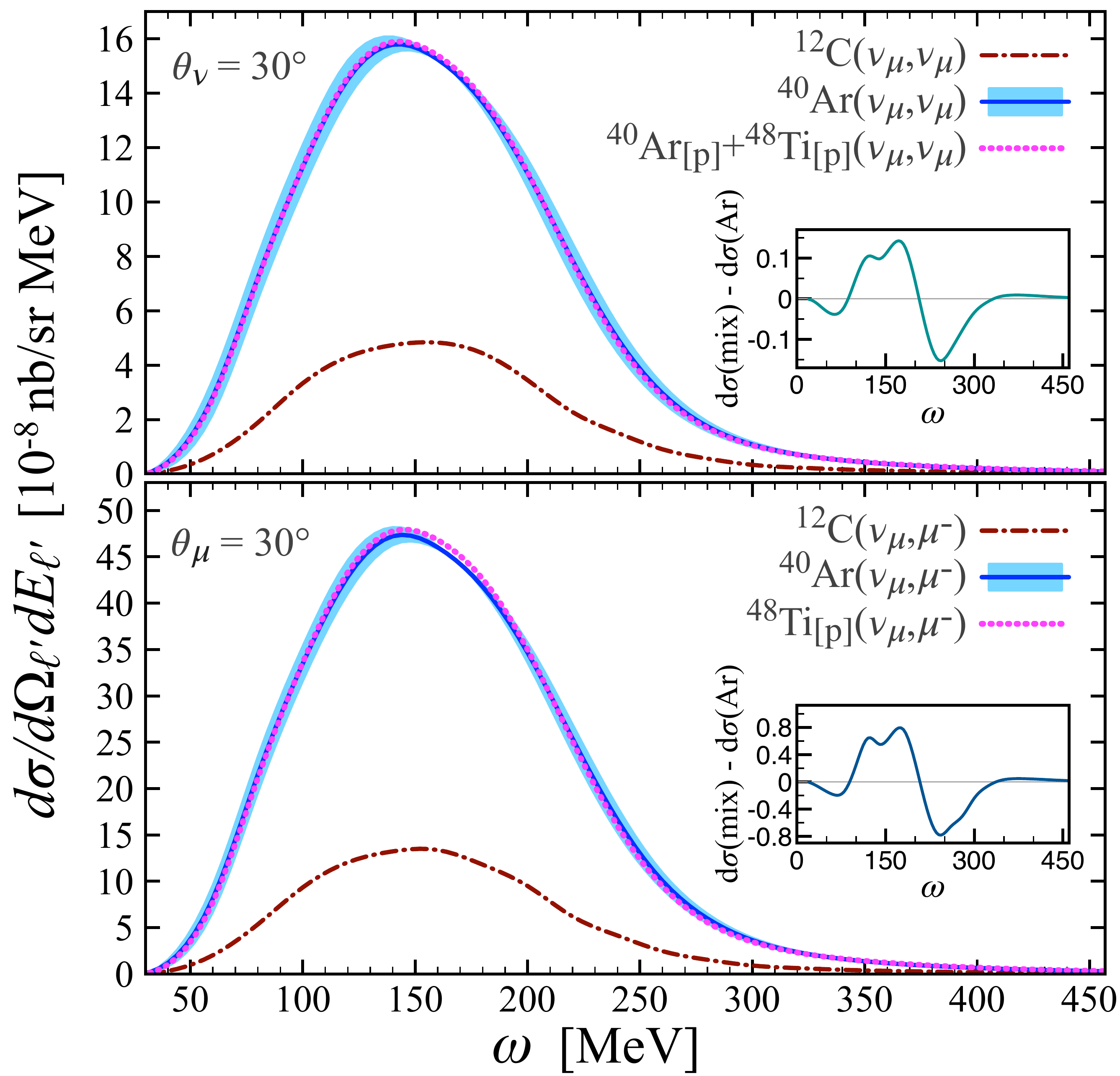}
\caption{Quasielastic neutral (top) and charged current (bottom) cross section for 1~GeV neutrino scattering. Dot-dashed lines refer to a $^{12}$C target and solid lines (with a color band showing the theoretical uncertainty due to model-space convergence) refer to $^{40}$Ar.
 The dotted lines result from using the $^{48}$Ti proton spectral function as an approximation for neutrons in $^{40}$Ar.
The insets show the difference between the latter and calculations where the full spectral distribution of $^{40}$Ar is used. \label{fig:nu_xsec} }
\end{figure}

Let us now turn to inclusive neutrino scattering on $^{40}$Ar, based on the SCGF spectral function and the reaction model discussed above. 
 The electroweak current is given by the sum of an axial and vector components. The latter is connected to the electromagnetic current through the conserved vector current hypothesis and is probed by electron scattering measurements. Figure~\ref{fig:nu_xsec} displays the computed inclusive cross sections at 1~GeV scattering energy for neutral and charged current reactions.  The dashed line shows the analogous calculation for $^{12}$C for comparison.  The quasielastic peak is found at similar transferred energies for both $^{40}$Ar and $^{12}$C
and its magnitude increases with the mass number, as expected from superscaling properties of inclusive reactions~\cite{Amaro2005Fy,Murphy2019JLab}.
 
  While in neutral current processes, the cross section depends on both the neutron and proton spectral functions, the charged current select only one of them. 
 In particular, charged current neutrino scattering probes the neutron spectral distribution of the nucleus. 
 The need to gain information on the neutron spectral distribution has indeed motivated the electron scattering measurements in Ti isotopes, whose proton number equals the neutron number of $^{40}$Ar, with the idea of exploiting isospin symmetry~\cite{Dai2018Tiee1}. 
 Besides the presence of the Coulomb potential, which results in an overall energy shift of the spectral function, it is not clear to which extent such a substitution is valid. 
 In particular, since the mirror isotope $^{40}$Ti is unstable and heavier Ti (mainly $^{48}$Ti) have to be used in electron scattering experiments, nuclear structure effects might play an important role.
To test the impact of this approximation we recomputed the cross sections of Fig.~\ref{fig:nu_xsec} substituting the neutron spectral function of $^{40}$Ar with the one computed for protons in $^{48}$Ti for both neutral and charged current processes.  The two curves result nearly identical at these energies, with discrepancies below 1\% (2\%) not only for neutral but also also for charged currents, where the validity of the replacement can be analyzed in greater detail. 

 \emph{Summary.} 
 We have computed the one-nucleon removal spectral functions of open-shell $^{40}$Ar and $^{48}$Ti isotopes, using \emph{ab initio} SCGF theory and saturating chiral interactions. Nuclear correlations were accounted for in the GGF-ADC(2) scheme that allows to obtain converged nuclear radii (with respect to many-body truncations) and crude yet quantitative predictions of the fragmentation of the spectral function.   The comparison with available electron scattering data are very satisfactory for both charge distributions and high-energy inclusive electron scattering up to the quasielastic peak.
 
 Based on this successful comparison, we used the spectral functions as input to predict inclusive neutrino cross sections on $^{40}$Ar at 1~GeV.  In this case, the quasielastic peak is centered at around missing energies of 150~MeV and extends up to $\simeq$~300~MeV. 
 Future studies will be needed (and will be possible within the present framework) to thoroughly assess all theoretical uncertainties, in particular those associated with the input Hamiltonian, ideally within a rigorous effective field theory approach.
Our findings support the hypothesis of Refs.~\cite{Dai2018Tiee1,Dai2019Ar40ee1} that approximating the mean-field neutron spectral distribution of $^{40}$Ar with the one for protons in Ti isotopes leads to very accurate results for neutrino scattering at the few GeV energies that are relevant to long-based neutrino oscillations experiments. 
 Further data from exclusive ($e$,$e$'$p$) measurements on $^{40}$Ar  and  Ti  will therefore be very important both to confront first-principle nuclear structure approaches and to constrain the  reaction rates needed for present and future generations of neutrino detectors.

{\em Acknowledgements.} 
This work has been supported by the Italian Centro Nazionale delle Ricerche (CNR) and the Royal Society under the CNR-Royal Society International Fellowship Scheme No. NF161046, 
by the United Kingdom Science and Technology Facilities Council (STFC) under Grants No. ST/P005314/1 and No. ST/L005816/1,
by the U.S. DOE, Office of Science, Office of Nuclear Physics, under contract DE-AC02-06CH11357
and by Fermi Research Alliance, LLC, under Contract No. DE-AC02-07CH11359 with the U.S. Department of Energy, Office of Science, Office of High Energy Physics.
Calculations were performed at the DiRAC DiAL system at the University of Leicester, UK, (BIS National E-infrastructure Capital Grant No. ST/K000373/1 and STFC Grant No. ST/K0003259/1), and using HPC resources from GENCI-TGCC, France, (Grant No. A005057392).
The work of NR has been supported by the NUclear Computational Low-Energy Initiative (NUCLEI) SciDAC project.

%\nocite{*}

\bibliography{Ar40_paper}% Produces the bibliography via BibTeX.

\end{document}